\documentclass[manuscript,screen]{acmart}
\AtBeginDocument{%
  \providecommand\BibTeX{{%
\normalfont B\kern-0.5em{\scshape i\kern-0.25em b}\kern-0.8em\TeX}}}

\setcopyright{rightsretained}
\copyrightyear{2024}
\acmYear{2024}
\acmDOI{}
\acmConference[CCAI 2024]{CHI 2024 Workshop on Child-centred AI Design}{May 11, 2024}{Honolulu, HI, USA}
\acmBooktitle{CHI 2024 Workshop on Child-centred AI Design, May 11, 2024, Honolulu, HI, USA}

\acmSubmissionID{AccessW17}

\usepackage{booktabs}
\usepackage{tabularx}
\usepackage{enumitem}
\usepackage{threeparttable}
\begin{document}

\title{Evaluation scheme for children-centered language interaction competence of AI-driven robots}

\author{Siqi Xie}
\authornote{Both authors contributed equally to this research.}
\email{sqxie@mails.ccnu.edu.cn}
\affiliation{%
  \institution{Central China Normal University}
  \streetaddress{152 Luoyu Rd.}
  \city{Wuhan}
  \state{Hubei}
  \country{China}
  \postcode{430079}
}
\orcid{0009-0007-7948-4560}

\author{Jiantao Li}
\authornotemark[1]
\email{lijt0909@163.com}
\affiliation{%
  \institution{Beijing Language and Culture University}
  \streetaddress{15 Xueyuan Rd.}
  \city{Beijing}
  \state{Beijing}
  \country{China}
  \postcode{100083}
}

\renewcommand{\shortauthors}{Siqi Xie \& Jiantao Li}

\begin{abstract}
This article explores the evaluation method for the language communication proficiency of AI-driven robots engaging in interactive communication with children. The utilization of AI-driven robots in children’s everyday communication is swiftly advancing, underscoring the importance of evaluating these robots’ language communication skills. Based on 11 Chinese families’ interviews and thematic analysis of the comment text from shopping websites (like Taobao, Tmall, and JD. et al.), a framework is introduced in the article to assess five key dimensions of child-robot language communication: interactivity, specificity, development, sociality, and safety. We draw on the concept of “children’s agency”, viewing children as active participants in shaping society and cultural life alongside adults. Therefore, this article places particular emphasis on collecting data related to children. Whether through survey interviews or direct interactive experiments, we treat children as an independent object for data collection. The study involved empirical research following the mentioned framework, which involved capturing interaction videos in natural conversation settings among children from 6 families. Analysis was performed on quantitative data obtained from video recordings, alongside questionnaires and interviews carried out by parents acting as participants or observers. We found that the presence or absence of parents during children’s interactions with robots can impact the evaluation of robots’ language communication abilities. Ultimately, this article proposes an enhanced comprehensive evaluation framework incorporating insights from parents and children, supported by empirical evidence and inter-rater consistency assessments, showcasing the scheme’s efficacy.
\end{abstract}




\maketitle

\section{Introduction}
The advent of artificial intelligence (AI) has ushered in a new era of technological advancement, significantly impacting various sectors including education and domestic environments. In recent years, the incorporation of AI-driven robots in child-oriented settings has witnessed substantial growth. These robots are no longer passive tools but active participants in fostering learning, social interaction, and emotional support for children \cite{grudin_tool_2017}. However, the effectiveness of these interactions largely hinges on the robot's ability to understand, process, and respond to children's language cues appropriately. Given children's unique linguistic characteristics and developmental stages, standard language interaction benchmarks used for adults may not be suitable. \par

Therefore, there is a pressing need for a dedicated evaluation scheme that comprehensively assesses a robot's language interaction competence specifically in the context of child users \cite{sun_exploring_2024}. To establish a comprehensive and adaptive evaluation scheme for children-centered language interaction competence, a mixed-methods approach was selected, which combines quantitative and qualitative methods to analyze and interpret the data,  including descriptive statistics, content analysis, and thematic analysis. Firstly, we have implemented a preliminary study to establish the evaluation framework, which interring 11 families and around 30 thousand words of comment test on shopping websites. Then six children between 3 and 6 years old were selected to interact with AI-driven robots, these children and their parents were consulted to develop the evaluation scheme based on a questionnaire and semi-structured interviews.\par
\section{Background and Related Work}
To evaluate the language interaction competence of robots, most research carried out on evaluating the effects of robot-assisted language learning \cite{belda-medina_using_2022, jeon_beyond_2023, kanero_social_2018, khalifa_learning_2019, randall_survey_2020, zhang_effect_2023},  emotion interaction \cite{catala_guidance_2023}or engagement degree\cite{de_haas_engagement_2022}, based on the efficiency-oriented approach. The components can be deduced from what kinds of functions or characters can improve the efficiency of HRI. For example, students evaluate robots in the robot-assisted language learning tasks from different aspects, including the understanding ability, speaking speed, sound quality, and English proficiency \cite{khalifa_learning_2019}; a conceptual framework consisting of three main components: goal orientation, embodiment, and multimodality was proposed to comprehensively understand different chatbot types and their possibilities for educational use in language learning \cite{jeon_beyond_2023}; Dautenhahn has proposed some dimensions for social robots \cite{dautenhahn_socially_2007}, but these kinds of dimensions are suitable for all the human beings, not specifically oriented to children, and not typically focused on social language skills.\par

In conclusion, these studies can’t comprehensively answer the following questions: What dimensions and components can affect the evaluation of child-directed language interaction competence of robots? How do family members affect child-robot interaction and the evaluation process? Despite the growing achievement, the research gaps are evident: recent research lacks theoretical frameworks to support the above questions and there is inadequate empirical research concerning the above questions.
\section{Preliminary Study}
The preliminary study aims to establish an outline of the evaluation framework, and we use the thematic analysis method to consult 11 Chinese families' views about children's language intelligence products. We collect data of semi-structured interviews and consumer feedback on online shopping platforms (such as Tmall\footnote{https://www.tmall.com} and JD\footnote{https://www.jd.com/}) to obtain information on the usage demands, experiences, and evaluations of children's robots (mainly in terms of interactive language). \par

\begin{figure}
    \centering
    \includegraphics[width=0.5\linewidth]{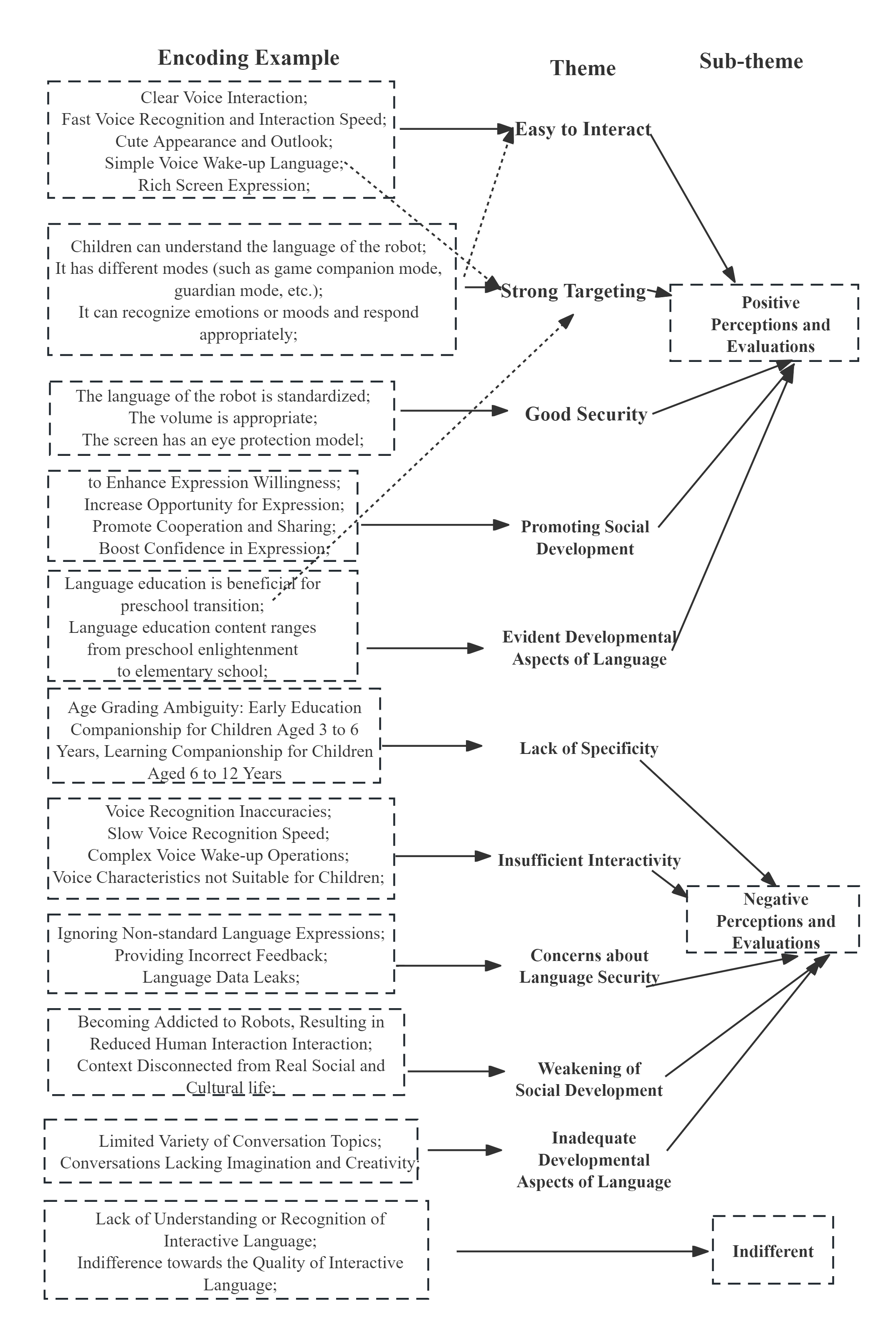}
    \caption{Thematic analysis}
    \label{fig: Thematic analysis}
    \Description{Thematic coding of the data}
\end{figure}

Following established open coding methods \citep{braun_using_2006}, the data obtained in this study was conducted by two researchers using thematic analysis. Firstly, the interviews were transcribed into text using Feishu Minutes software\footnote{https://www.feishu.cn/product/minutes}, and manual verification was conducted to enhance the understanding of the entire interview content. After transcription, the data were read multiple times to develop initial coding awareness. Then, initial coding was conducted by two coders independently analyzing the interview data, comparing similarities and differences, and using Nvivo for inductive and theory-driven thematic analysis. This approach allowed for the discovery of facts based on the interview data and provided specific investigations into the interactive language of children's robots. Next, themes were identified by comparing and categorizing the initial codes, grouping conceptually similar codes, and assigning different codes to different themes. Only the initial codes agreed upon by both coders were confirmed as final codes. Finally, the themes were examined to ensure the accuracy of the coding and the relevance to the respective theme. The specific theme codes can be found in Fig.\ref{fig: Thematic analysis}.

Furthermore, a saturation test was conducted using 411 feedback evaluations from shopping platforms. The analysis results showed that during the coding analysis process, the identified codes and themes were repeatedly present, and no new codes or themes emerged. Therefore, through the saturation test, it can be concluded that the extracted codes and themes in this study are reliable.
\section{Empirical Research}
Only when observing the real interactive context can different evaluation dimensions be tested, in this case, a comprehensive evaluation framework of robots' children-centered language ability will be proposed. Our research's objective is to explore whether the five dimensions of the evaluation framework can comprehensively assess the language interaction competence of robots.
\subsection{Methods}
The research is centered on preschool-aged children due to their higher likelihood of being cared for at home compared to older children. A total of 6 kindergartens between 3 and 6 years old and their parents from China participated in this study. And the study involved parental participants from two age ranges: 20-30 years and 30-40 years, with three parents in each group. They also had a range of education levels from bachelor to doctoral degrees. \par
We use a robot called Alpha Egg GPT, which is LLM-equipped for managing the dialogue flow (i.e., The robot was embedded in the API of iFlytek's Alpha-egg large cognitive model for children). Parents were requested to complete an online survey, providing demographic information about their family and their previous encounters with robots. Subsequently, the researcher arranged experiments with the parents and their child, either at their residence or within a laboratory setting. The engagement between the individual child and the robots extended for approximately 30 minutes, encompassing a series of prescribed procedures  followed by Table\ref{tab:procedure}:
\begin{table}[t]\small
  \centering
\caption{Procedure of empirical child-robot interaction}
\label{tab:procedure}
\begin{tabularx}{\columnwidth}{llX}
\toprule
Setting & Stage & Description \\ \midrule
Parents are absent& Ice-breaking & Introduction of their name, age, and hobbies.\\
& Explanation & Tutorial presentation on how to use the robot.\\
& Open-dialogue & Children interact with the robot in an open-domain mode, such as asking questions, sharing experiences, or storytelling, and the investigators observe them during the entire process.\\
Parents are present& Introduction & Introduction of the parent's name, and to establish the identity of the participant.\\ 
& Collaboration & Parents and children co-create a story or share a series of emotional experiences during children's study and life.\\ \bottomrule
\end{tabularx}
\end{table}
Throughout our study, we meticulously recorded all interaction sessions between children and robots. These video recordings serve as a valuable resource, capturing the nuances of behavior, communication, and emotional responses. By analyzing these videos, we gain insights into how children engage with robots. Non-verbal cues, gestures, and expressions provide valuable context that might not be apparent through other data collection methods.\par
In addition to video recordings, we conduct post-interviews with both parents and children. These interviews provide subjective perspectives and rich contextual information. Following the practical session of each family, a brief 10-minute interview was conducted, primarily aimed at gathering insights and experiences from both parents and children regarding their feelings in the respective session. After observing and receiving participants' feedback, we have developed a comprehensive evaluation scheme based on 5 dimensions extracted from our preliminary study. More details will be provided in the following section.\par
We transcribed videos by Baidu Netdisk\footnote{https://pan.baidu.com} and noted the timestamps of positive and negative feedback from children. Then, we asked parents and children to complete a questionnaire based on the developed evaluation scheme. This questionnaire reports the degree of satisfaction, which is a quantitative measure used to assess the success of child-robot interactions.
\subsection{Developed scheme}
The preliminary research involved the implementation of an extensive assessment framework. We established a detailed evaluation scheme based on empirical research. The evaluation system, outlined in Table \ref{Dimensions}, encompasses a total of five dimensions and 16 indicators.
\begin{table}[t]
\renewcommand\arraystretch{1}
    \centering
    \caption{Developed evaluation scheme}
    \resizebox{\linewidth}{!}{
    \begin{tabularx}{1\columnwidth}{lX}
    \toprule
        \textbf{Dimensions} & \textbf{Indicators} \\ \midrule
        {Interactivity} & $\bullet$ Response are relevant to the context, purpose and tasks, and feedback can stimulate children's desire to further express themselves.  \\ 
        {} & $\bullet$ Speech recognition is accurate, and the language output or response is smooth and rapid. \\ 
        {} & $\bullet$ Can adjust the style of dialogue, content, and strategy according to children's feedback, prompt and the whole family's collaboration. \\ 
        {Specificity} & $\bullet$ The voice features meet the needs of children and have affinity, children can customize the voice role;\\ 
        {} & $\bullet$ Language interaction is in line with children's cognition, emotions, and interests, with empathy and fun; \\ 
        {} & $\bullet$ The vocabulary and syntax of the language output meet the age level of the children. \\ 
        {Expansibility} & $\bullet$ The theme of the session was appropriately challenging to the children's level of cognitive development; \\     
        {} & $\bullet$ Adjusting for language interaction content and communicative strategies with children's age and cognitive development; \\ 
        {} &$\bullet$  Can update the system repository or language model version, and enrich the children-centered language resources. \\ 
        {Safety} & $\bullet$ Filter out false, unverified, untrue or other inappropriate material; \\ 
        {} & $\bullet$ Refusing to answer and export contents that imply negative energy, violence, or discrimination that affect children's physical and mental health; \\ 
        {} & $\bullet$ Can protect privacy and data security; \\ 
        {} & $\bullet$ Can interrupt dialogue and self-lock to prevent children from becoming too addicted. \\ 
        {Sociability} & $\bullet$ Language output conforms to social culture life, law, moral and language norms; \\ 
        {} & $\bullet$ Being able to promptly correct and guide children's speech behaviors that do not comply with moral and language norms;\\ 
        {} & $\bullet$ Interpreting children's social intent correctly and engaging with them appropriately like a real social role.\\ 
        \bottomrule
    \end{tabularx}}
    \label{tab:Dimensions}
\end{table}

\subsection{Result and Findings}
To evaluate the efficiency of the developed evaluation scheme, parents finished a questionnaire and made a 5-rate scale to score the different dimensions. They are asked to assess it by the practical experience and the video recording of the interaction. In the end, we calculate the mean score, SD, and internal reliability of each dimension, Table \ref{tab: evaluation} shows the result. The Cronbach $\alpha$ score serves as an essential metric for assessing the reliability of the developed evaluation scheme.\par
\begin{table}[t]
    \centering
    \caption{Evaluation result}
    \begin{threeparttable}
    \begin{tabular}{lcc}
    \toprule
        {Dimensions} & {Mean±SD} & {Internal reliability\tnote{1}}\\ \midrule
        Interactivity & 1.778±0.344 & .750 \\ 
        Specificity & 2.278±0.443 & .906 \\ 
        Expansibility & 2.778±0.344 & .750 \\ 
        Safety & 3.292±0.332 & .805 \\ 
        Sociality & 2.722±0.390 & .732 \\ \bottomrule
    \end{tabular}
    \begin{tablenotes} 
        \footnotesize
        \item[1] The internal reliability used Cronbach $\alpha$ to calculate
    \end{tablenotes} 
    \end{threeparttable} 
    \label{tab: evaluation}
\end{table}

The study found that the complex communication style of robots, which included the use of sophisticated vocabulary and intricate grammatical structures, proved to be a hurdle for the children involved. Simplifying language, using more common words and shorter sentences, is crucial to enhance the effectiveness of robot-child interactions. Secondly, parental involvement plays a crucial role in fostering a child's willingness to engage and participate actively, and supportive environments contribute to a more positive and immersive experience for the child. Thirdly, the robots exhibited significant limitations when it came to discerning individual voice characteristics amidst a cacophony of multiple speakers, underscoring a pressing need for the enhancement of multi-part voice recognition capabilities.
\section{Discussion}
\paragraph{Balance of Safety and Creativity} 
The delicate equilibrium between safety and creativity lies at the heart of designing effective language interaction competence evaluation schemes for AI-driven robots. While fostering creativity is essential for engaging children and promoting cognitive development, ensuring safety remains paramount. Striking this balance necessitates thoughtful consideration of the following aspects:
1. Content Filters: Implementing robust content filters to shield children from harmful or inappropriate material while still allowing for imaginative and educational interactions.
2. Dynamic Adaptation: Developing adaptable algorithms that adjust their responses based on context, age, and individual preferences, thereby fostering creativity without compromising safety.
3. Human Oversight: Regular human monitoring and intervention to address unforeseen situations and maintain a safe environment.
\paragraph{Considering the Role of AI-Driven Robots}
Recognizing the multifaceted roles AI-driven robots play in children's lives is crucial. Beyond mere communication tools, they serve as companions, educators, and playmates.
1. Educational Enhancement: Leveraging robots to enhance language learning, cognitive skills, and social development.
2. Emotional Support: Acknowledging the potential for robots to provide emotional support and companionship, especially in contexts where human interaction is limited.
3. Ethical Responsibility: Ensuring that robots uphold ethical standards and align with societal values, considering their influential role in shaping children's perceptions and behaviors.
\paragraph{Family Members' Participation}
Involving family members—particularly parents and guardians—in the evaluation process is pivotal. Their active participation contributes valuable insights and fosters a holistic understanding of child-robot interactions.
1. Parental Perception: Recognizing that parents' perspectives significantly impact the evaluation of robots' language communication abilities.
2. Collaborative Assessment: Encouraging joint assessments by parents and children, as their combined viewpoints provide a comprehensive picture.
3. Feedback Loop: Establishing a continuous feedback loop between families, researchers, and developers to refine and improve language interaction competence.
In summary, thoughtful consideration of safety, the multifaceted role of AI-driven robots, and family involvement will shape effective evaluation schemes, ensuring that child-robot language interactions are not only proficient but also enriching and secure.
\section{Conclusion}
We present a conceptual model aimed at gauging the linguistic capabilities of AI-powered robots engaging with children. Our scheme scrutinizes five facets of communication between children and robots: responsiveness, precision, progression, sociability, and security. Information was gathered from six family units via video documentation, surveys, and personal discussions. Our results underscore the necessity for a thorough assessment protocol integrating perspectives from both parental figures and children.

\section{Acknowledgments}
We express our great gratitude to all the participating families, whose time and efforts were essential to the success of this study. Our special gratitude is extended to participant Zhenda Zhang and his son Dudu, for their remarkable dedication and invaluable contribution to this research. Furthermore, we would like to acknowledge the support provided by the National Social Science Foundation of China (Major Program) (23\&ZD320) and the Fundamental Research Funds for the Central Universities (No. 30106230479).




\bibliographystyle{ACM-Reference-Format}
\bibliography{references}


\end{document}